\documentclass[useAMS,usenatbib,referee]{biom} 
\def\bSig\mathbf{\Sigma}

\usepackage{amsmath}
\usepackage{psfrag,epsf}
\usepackage{enumerate}
\usepackage{url} 
\usepackage[export]{adjustbox}
\usepackage{subfig}
\usepackage{float}
\usepackage{color}
\usepackage{tikz}
\usetikzlibrary{positioning}
\usepackage{standalone}
\usepackage{algpseudocode}
\usepackage{algorithm}
\usepackage[toc,title,page]{appendix}
\usepackage{amsfonts}

\newcommand{\bx}{\boldsymbol}

\title[Efficient Computation of High-Dimensional Penalized Generalized Linear Mixed Models]{Efficient Computation of High-Dimensional Penalized Generalized Linear Mixed Models by Latent Factor Modeling of the Random Effects}

\author{Hillary M. Heiling$^{1,*}$\email{hmheiling@gmail.com}, 
Naim U. Rashid$^{1}$,
Quefeng Li$^{1}$,
Xianlu L. Peng$^{2}$,
Jen Jen Yeh$^{2,3,4}$,\\
\textbf{and 
Joseph G. Ibrahim$^{1}$} \\ 
$^{1}$Department of Biostatistics, University of North Carolina Chapel Hill, Chapel Hill, NC \\
$^{2}$Lineberger Comprehensive Cancer Center, University of North Carolina at 
Chapel Hill, Chapel Hill, NC \\
$^{3}$Department of Surgery, University of North Carolina Chapel Hill, Chapel Hill, NC \\
$^{4}$Department of Pharmacology, University of North Carolina Chapel Hill, Chapel Hill, NC}

\begin{document}


\date{2023}



\pagerange{\pageref{firstpage}--\pageref{lastpage}} 




\label{firstpage}


\begin{abstract}
Modern biomedical datasets are increasingly high dimensional and exhibit complex correlation structures. Generalized Linear Mixed Models (GLMMs) have long been employed to account for such dependencies. However, proper specification of the fixed and random effects in GLMMs is increasingly difficult in high dimensions, and computational complexity grows with increasing dimension of the random effects. We present a novel reformulation of the GLMM using a factor model decomposition of the random effects, enabling scalable computation of GLMMs in high dimensions by reducing the latent space from a large number of random effects to a smaller set of latent factors. We also extend our prior work to estimate model parameters using a modified Monte Carlo Expectation Conditional Minimization algorithm, allowing us to perform variable selection on both the fixed and random effects simultaneously. We show through simulation that through this factor model decomposition, our method can fit high dimensional penalized GLMMs faster than comparable methods and more easily scale to larger dimensions not previously seen in existing approaches. 
\end{abstract}

%



\maketitle


%

\section{Introduction}
\label{sec:intro}


Modern biomedical datasets are increasingly high dimensional and exhibit complex correlation structures.
Generalized Linear Mixed Models (GLMMs) have long been employed to account for such dependencies. However, proper specification of the fixed and random effects is a critical step in estimation of GLMMs. For instance, omitting important random effects can lead to bias in the estimated variance of the fixed effects \citep{gurka2011avoiding, bondell2010joint}, whereas including unnecessary random effects could increase the computational difficulty of fitting the GLMM. 
Despite the importance of properly specifying the set of fixed and random effects in such models, it is often unknown \textit{a priori} which variables should be specified as fixed or random in the model, particularly in high dimensional settings in which the feature space of both the fixed and random effects is generally assumed to be sparse.

There are several existing methods used to select fixed and/or random effects within mixed models. One such class of methods are what we term here as candidate model selection. In these methods, researchers manually specify a set of candidate models, use existing software such as R packages \textbf{lme4} \citep{lme42007} 
or \textbf{MCMCglmm} \citep{mcmcglmm2010} to fit the candidate models, and then select the best one using appropriate mixed effects model selection criteria such as the BIC-ICQ criterion \citep{BICq2011}, the hybrid Bayesian information criterion BICh \citep{BICh2014}, or fence methods \citep{jiang2014fence}. 
However, candidate model selection approaches are not feasible for moderate or large dimensions as there are $2^{2p}$ possible combinations of fixed and random effects for $p$ predictors of interest. 

Penalized likelihood techniques, such as LASSO, have been expanded to apply to mixed models. However, most of these existing approaches have limitations in their applicability. Some methods only select fixed effects, such as the R packages \textbf{glmmLasso} \citep{groll2014variable} 
and \textbf{glmmixedLASSO} \citep{schelldorfer2014glmmlasso}, or only select random effects \citep{pan2014random}. Other methods that do select both fixed and random effects are generally limited in their scalability due to their computational burden in high dimensions. Examples include the adaptive Lasso method proposed by \cite{BICq2011}, which utilizes a computationally intensive expectation step in their Monte Carlo Expectation Minimization (MCEM) algorithm, and the regularized PQL method proposed by \cite{hui2017joint}, which has prohibitive initialization requirements in high dimensions and requires the calculation of an inverse matrix with dimensions equal to the random effects remaining in the model. 

\cite{rashid2020} developed a penalization method that selects both fixed and random effects in larger dimensions not seen in previous approaches.
The method developed by \cite{rashid2020}, which uses a Monte Carlo Expectation Conditional Minimization (MCECM) algorithm, was applied to simulations and a case study comprised of 50 covariates and has since been incorporated into the \textbf{glmmPen} R package \citep{glmmPen_pkg}.
Although this \textbf{glmmPen} framework extends the feasible dimensionality of performing variable selection within GLMMs relative to existing methods, new methodology is needed to alleviate the computational burden as the dimension increases even further.

We present a novel reformulation of the GLMM, which we call \textbf{glmmPen\_FA}, using a factor model decomposition of the random effects. This factor model is used as a dimension reduction tool to represent a large number of latent random effects as a function of a smaller set of latent factors. By reducing the latent space of the random effects, this new model formulation enables us to extend the feasible dimensionality of performing variable selection in GLMMs to hundreds of predictors. We estimate model parameters and perform simultaneous selection of fixed and random effects using
an MCECM algorithm.
We show through simulations that through this factor model decomposition, our method can fit high dimensional penalized GLMMs (pGLMMs) faster than comparable methods and more easily scale to larger dimensions not previously seen in existing approaches.

We illustrate the utility of our method by applying our method to a case study that we present in Section 4. The case study data combines gene expression data from pancreatic ductal adenocardinoma patients across five separate studies. 
We aim to select important features that predict the pancreatic cancer subtypes basal or classical \citep{moffitt2015} (i.e. identify non-zero fixed effects) as well as identify features that have varied effects across the groups (i.e. identify non-zero random effects). Due to the large number of features that we consider, it is difficult to have \textit{a priori} knowledge of which features have non-zero fixed or random effects. Therefore, we will use our method to fit a penalized logistic mixed effects model to select important fixed and random effects.

The remainder of this paper is organized as follows. 
Section \ref{sec:methods} reviews the statistical models and algorithm used to estimate pGLMMs in our new factor model decomposition framework, termed \textbf{glmmPen\_FA}. In Section \ref{sec:sims}, simulations are conducted to assess the performance of the new \textbf{glmmPen\_FA} method. 
Section \ref{sec:PDAC} describes the motivating case study, where we aim to utilize our method to identify gene expression features important in the prediction of pancreatic cancer subtypes.
We close the article with some discussion in Section \ref{sec:discuss}.

\section{Methods}
\label{sec:methods}

\subsection{Model formulation}
\label{sec:models}

In this section, we review the notation and model formulation of our approach.
We consider the case where we want to analyze data from \(K\) independent groups of observations. For each group \(k = 1,...,K\), there are \(n_k\) observations for a total sample size
of \(N = \sum_{k=1}^K n_k\). 
For group $k$, let
\(\boldsymbol y_k = (y_{k1},...,y_{kn_k})^T\) be the vector of \(n_k\)
independent responses, \(\boldsymbol x_{ki} = (x_{ki,1},...,x_{ki,p})^T\) be the
\(p\)-dimensional vector of predictors, and \(\boldsymbol X_k = (\boldsymbol x_{k1}, ..., \boldsymbol x_{kn_k})^T\).
For simplification of notation, we will set $n_1 = ... = n_K = n$ without loss of generality. In GLMMs, we
assume that the conditional distribution of \(\boldsymbol y_k\) given
\(\boldsymbol X_k\) belongs to the exponential family and has the
following density: \begin{equation}
  f(\boldsymbol y_k | \boldsymbol X_k, \boldsymbol \alpha_k; \theta) = 
    \prod_{i=1}^{n} c(y_{ki}) \exp[\tau^{-1} \{y_{ki} \eta_{ki} - b(\eta_{ki})\}],
    \label{eqn:chp4_exp_family}
\end{equation} where \(c(y_{ki})\) is a constant that only depends on
\(y_{ki}\), \(\tau\) is the dispersion parameter, \(b(\cdot)\) is a known
link function, \(\eta_{ki}\) is the linear predictor, $\boldsymbol{\alpha_k}$ are group-specific latent variables, and $\boldsymbol{\theta}$ represents all model coefficients (see more detailed descriptions later in this section). We standardize the fixed effects covariates matrix
\(\boldsymbol X = (\boldsymbol X_1^T,...,\boldsymbol X_K^T)^T\) such that \\ \(\sum_{k=1}^K \sum_{i=1}^{n_{k}} x_{ki,j} = 0\) and
\(N^{-1} \sum_{k=1}^K \sum_{i=1}^{n_{k}} x_{ki,j}^2 = 1\) for
\(j = 1,...,p\). 

As outlined in \cite{rashid2020}, the traditional GLMM formulation of the linear predictor can be represented as
\begin{equation}
  \eta_{ki} = \boldsymbol x_{ki}^T \boldsymbol \beta + \boldsymbol z_{ki}^T \boldsymbol \gamma_k = \boldsymbol x_{ki}^T \boldsymbol \beta + \boldsymbol z_{ki}^T \boldsymbol \Gamma \boldsymbol \delta_k,
  \label{eqn:chp4_linpredgam}
\end{equation} where \(\boldsymbol \beta = (\beta_1,...,\beta_{p})^T\)
is a \(p\)-dimensional vector for the fixed effects coefficients (including the intercept), \(\boldsymbol z_{ki}\) is a \(q\)-dimensional
subvector of \(\boldsymbol x_{ki}\) representing the random effect predictors (including the random intercept), $\boldsymbol \Gamma$ is the Cholesky decomposition of the random effects covariance matrix $\boldsymbol \Sigma$ such that $\boldsymbol \Gamma \boldsymbol \Gamma^T = \boldsymbol \Sigma$, and $\boldsymbol \gamma_k = \boldsymbol \Gamma \boldsymbol \delta_k$ is a $q$-dimensional vector of unobservable random effects for group $k$ where $\boldsymbol \delta_k \sim N_q(0,I)$. 


In its current representation, the model assumes a latent space of dimension $q$, the number of random effect predictors.
When $q$ is large, the estimation of the covariance matrix $\boldsymbol \Sigma$ = Var($\boldsymbol \gamma_k$) can be computationally burdensome to compute due to both the number of parameters needed to estimate this matrix ($q(q+1)/2$ parameters are needed for an unstructured covariance matrix) as well as the need to approximate a $q$-dimensional integral (see Section \ref{sec:model_advantages} for details). 
Prior work such as \cite{fan2013POET} and \cite{tran2020bayesian} have assumed a factor model structure in order to estimate high-dimensional covariance matrices in other settings, such as the estimation of sample covariance matrices for time series data and the covariance matrix in variational inference used to approximate the posterior distribution, respectively. Here we introduce a novel formulation of the GLMM where we decompose the random effects $\boldsymbol \gamma_k$ into a factor model with $r$ latent common factors ($r \ll q$) such that $\boldsymbol \gamma_k = \boldsymbol B \boldsymbol \alpha_k$, where $\boldsymbol B$ is the $q \times r$ loading matrix and $\boldsymbol \alpha_k$ represents the $r$ latent common factors. We assume the latent factors $\boldsymbol \alpha_k$ are uncorrelated and follow a $N_r(\boldsymbol 0, \boldsymbol I)$ distribution. We re-write the linear predictor as
\begin{equation}
  \eta_{ki} = \boldsymbol x_{ki}^T \boldsymbol \beta + \boldsymbol z_{ki}^T \boldsymbol B \boldsymbol \alpha_k.
  \label{eqn:chp4_linpredB}
\end{equation}
In this representation, the random component of the linear predictor has variance Var(\(\boldsymbol B \boldsymbol \alpha_k\)) = \(\boldsymbol{B B}^T\) = $\boldsymbol \Sigma$. By assuming that $\boldsymbol \Sigma$ is low rank, we also reduce the dimension of the latent space from $q$ to $r$, which reduces the dimension of the integral in the likelihood and thereby reduces the computational complexity of the E-step in the EM algorithm. Further details are given in Section \ref{sec:model_advantages}.

In order to estimate \(\boldsymbol B\), let $\boldsymbol b_t \in \mathbb{R}^r$ be the $t$-th row of $\boldsymbol B$ and $\boldsymbol b = (\boldsymbol b_1^T,...,\boldsymbol b_q^T)^T$. 
We can then reparameterize the linear predictor as
\begin{equation}
  \eta_{ki} = \boldsymbol x_{ki}^T \boldsymbol\beta + \boldsymbol z_{ki}^T \boldsymbol B \boldsymbol\alpha_k = \left (\boldsymbol x_{ki}^T \hspace{10 pt} (\boldsymbol\alpha_k \otimes \boldsymbol z_{ki})^T \boldsymbol J \right)
  \left (
  \begin{matrix}
    \boldsymbol\beta \\ \boldsymbol b 
  \end{matrix}
  \right )
\end{equation} in a manner similar to \cite{chen2003} and \cite{BICq2011}, where \(\boldsymbol J\) is a matrix that transforms $\boldsymbol b$ to vec($\boldsymbol B$) such that $\text{vec}(\boldsymbol B) = \boldsymbol J \boldsymbol b$ and \(\boldsymbol J\) is of
dimension $(qr)\times(qr)$. The vector of parameters
\(\boldsymbol \theta = (\boldsymbol \beta^T, \boldsymbol b^T, \tau)^T\)
are the main parameters of interest. We denote the true value of
\(\boldsymbol \theta\) as
\(\boldsymbol \theta^{*} = (\boldsymbol \beta^{*T}, \boldsymbol b^{*T}, \tau^{*})^T = \text{argmin}_{\boldsymbol \theta}\text{E}_{\boldsymbol\theta}[-\ell(\boldsymbol \theta)]\)
where \(\ell(\boldsymbol \theta)\) is the observed
log-likelihood across all \(K\) groups such that
\(\ell(\boldsymbol \theta) = \sum_{k=1}^K \ell_k(\boldsymbol \theta)\), where
\(\ell_k(\boldsymbol \theta) = (1/n) \log \int f(\boldsymbol y_k | \boldsymbol X_k, \boldsymbol \alpha_k; \boldsymbol \theta) \phi(\boldsymbol \alpha_k) d \boldsymbol \alpha_k\).

Our main interest lies in selecting the
true nonzero fixed and random effects. In other words, we
aim to identify the set $S = S_1 \cup S_2 = \{j: \beta_j^* \ne 0 \} \cup \{t: ||\boldsymbol b_t^*||_2 \ne 0\}$,
where \(S_1\) and \(S_2\) represent the true fixed and random effects, respectively. When \(\boldsymbol b_t = \boldsymbol 0\),
this indicates that the effect of covariate \(t\) is fixed across the \(K\)
groups (i.e. the corresponding $t$-th row and column of $\boldsymbol \Sigma$ is set to $\boldsymbol 0$).

We aim to solve the following penalized likelihood problem: \begin{equation}
  \widehat{\boldsymbol\theta} = \text{argmin}_{\boldsymbol\theta} - \ell (\boldsymbol\theta) + \lambda_0 \sum_{j=1}^{p} \rho_0 \left (\beta_j \right ) + \lambda_1 \sum_{t=1}^{q} \rho_1 \left (||\boldsymbol b_t||_2 \right ),
  \label{eqn:chp4_penlik}
\end{equation} where \(\ell(\boldsymbol \theta)\) is the observed log-likelihood for all \(K\) groups defined earlier,
\(\rho_0(t)\) and \(\rho_1(t)\) are general folded-concave penalty
functions, and \(\lambda_0\) and \(\lambda_1\) are positive tuning
parameters. The \(\rho_0(t)\) penalty
functions could include the \(L_1\) penalty, the SCAD penalty,
and the MCP penalty \citep{glmnet2010, ncvreg2011}. For the
\(\rho_1(t)\) penalty, we treat the elements of $\boldsymbol b_t$ as a group and penalize them in a groupwise manner using the group
LASSO, group MCP, or group SCAD penalties presented by Breheny and Huang
\citeyearpar{grpreg2015}. These groups of $\boldsymbol b_t$ are
then estimated to be either all zero or all nonzero. In this way, we
select covariates to have varying effects
(\(\boldsymbol{\widehat b}_t \ne \boldsymbol 0\)) or fixed effects
(\(\boldsymbol{\widehat b}_t = \boldsymbol 0\)) across the \(K\)
groups.


\subsection{MCECM algorithm}
\label{sec:chp4_mcecm}

We solve (\ref{eqn:chp4_penlik}) for some specific penalty combination $(\lambda_0,\lambda_1)$ using a Monte Carlo Expectation
Conditional Minimization (MCECM) algorithm \citep{garcia2010}. 

In the \(s^{th}\) iteration of the MCECM algorithm, we aim to evaluate the expectation of (E-step) and minimize (M-step) the following penalized Q-function: \begin{align}
  \begin{aligned}
    Q_\lambda(\boldsymbol\theta | \boldsymbol \theta^{(s)}) & = \sum_{k=1}^K E \left \{ -\log(f(\boldsymbol y_k, \boldsymbol X_k, \boldsymbol\alpha_k;\boldsymbol\theta | \boldsymbol d_o; \boldsymbol\theta^{(s)})) \right \} + \\
    & \hspace{10pt} \lambda_0 \sum_{j=1}^{p} \rho_0 \left (\beta_j \right ) + \lambda_1 \sum_{t=1}^{q} \rho_1 \left (||\boldsymbol b_t||_2 \right ) \\
    & = Q_1(\boldsymbol\theta | \boldsymbol\theta^{(s)}) + Q_2(\boldsymbol\theta^{(s)}) + \lambda_0 \sum_{j=1}^{p} \rho_0 \left (\beta_j \right ) + \lambda_1 \sum_{t=1}^{q} \rho_1 \left (||\boldsymbol b_t||_2 \right ),
    \label{eqn:chp4_Qfun}
  \end{aligned}
\end{align} where
\((\boldsymbol y_k, \boldsymbol X_k, \boldsymbol \alpha_k)\) gives the
complete data for group $k$,
\(d_{k,o} = (\boldsymbol y_k, \boldsymbol X_k)\) gives the observed data
for group $k$, \(\boldsymbol d_o\) represents the entirety of the observed
data, and \begin{equation}
  Q_1(\boldsymbol\theta | \boldsymbol\theta^{(s)}) = - \sum_{k=1}^K \int \log [f(\boldsymbol y_k | \boldsymbol X_k, \boldsymbol\alpha_k; \boldsymbol\theta)] \phi(\boldsymbol\alpha_k | \boldsymbol d_{k,o}; \boldsymbol\theta^{(s)}) d \boldsymbol\alpha_k,
  \label{eqn:chp4_Q1}
\end{equation} \begin{equation}
   Q_2(\boldsymbol\theta^{(s)}) = - \sum_{k=1}^K \int \log [\phi(\boldsymbol \alpha_k)] \phi(\boldsymbol\alpha_k | \boldsymbol d_{k,o}; \boldsymbol\theta^{(s)}) d \boldsymbol\alpha_k
  \label{eqn:chp4_Q2}
\end{equation}


\subsubsection{Monte-Carlo E-step} 
The E-step of the algorithm aims to approximate the $r$-dimensional integral expressed in (\ref{eqn:chp4_Q1}).
The integrals in the Q-function do not have closed forms when $f(\boldsymbol y_k | \boldsymbol X_k, \boldsymbol\alpha_k^{(s,m)}; \boldsymbol\theta)$ is assumed to be non-Gaussian. We approximate these integrals using a Markov
Chain Monte Carlo (MCMC) sample of size M from the posterior density
\(\phi(\boldsymbol \alpha_k | \boldsymbol d_{k,o}; \boldsymbol \theta^{(s)})\). Let \(\boldsymbol \alpha_k^{(s,m)}\) be the
\(m^{th}\) simulated $r$-dimensional vector from the posterior of the latent common factors, \(m = 1,...,M\), at the \(s^{th}\) iteration
of the algorithm for group \(k\). The integral in (\ref{eqn:chp4_Q1}) can be approximated as \begin{equation}
  Q_1(\boldsymbol\theta | \boldsymbol\theta^{(s)}) \approx - \frac{1}{M} \sum_{m=1}^M \sum_{k=1}^K \log f(\boldsymbol y_k | \boldsymbol X_k, \boldsymbol\alpha_k^{(s,m)}; \boldsymbol\theta).
  \label{eqn:qfun_approx}
\end{equation}
We use the fast and efficient No-U-Turn Hamiltonian Monte Carlo (NUTS HMC) sampling procedure from the Stan software \citep{stan2017}
in order to perform the E-step efficiently.

\subsubsection{M-step}
\label{sec:chp4_mstep}
In the M-step of the algorithm, we aim to minimize \begin{equation}
  Q_{1,\lambda}(\boldsymbol\theta | \boldsymbol\theta^{(s)}) = Q_1(\boldsymbol\theta | \boldsymbol\theta^{(s)}) + \lambda_0 \sum_{j=1}^{p} \rho_0 \left (\beta_j \right ) + \lambda_1 \sum_{t=1}^{q} \rho_1 \left (||\boldsymbol b_t||_2 \right )
  \label{eqn:chp4_Mstep}
\end{equation} with respect to
\(\boldsymbol \theta = (\boldsymbol \beta^T, \boldsymbol b^T, \tau)^T\). We do this by using a Majorization-Minimization algorithm with penalties applied to the fixed effects and the rows of $\boldsymbol B$.
The M-step of the \(s^{th}\) iteration of the MCECM algorithm proceeds as described in Algorithm 1 given in Web Appendix Section 1.1. 

\subsubsection{MCECM algorithm} 

Algorithm 2 in Web Appendix Section 1.1
describes the full MCECM algorithm for estimating the parameters with a particular penalty combination $(\lambda_0,\lambda_1)$. The process of model selection and finding optimal tuning parameters are described in full in the Web Appendix (see Sections 1.2 and 1.3). Briefly, we identify a maximum penalty that penalizes all fixed effects to zero using techniques from the \textbf{ncvreg} R package \citep{ncvreg2011}. We calculate a sequence of penalties from a small proportion of the max penalty (the minimum penalty) to the max penalty that are equidistant on the log scale, and apply this sequence to both the fixed and random effects in a two-stage model selection approach.
For further details on initialization and convergence, also see the Web Appendix Section 1.4.




\subsection{Advantages of \textbf{glmmPen\_FA} model formulation}
\label{sec:model_advantages}

There are several advantages to our proposed factor model decomposition of the random effects. By representing the random effects with a factor model, we reduce the latent space from a high dimension of $q$ (the number of candidate random effect predictors) to $r$.
In the more traditional GLMM model formulation, $Q_1(\boldsymbol \theta | \boldsymbol \theta^{(s)})$ would be represented as 
\begin{equation}
  Q_1(\boldsymbol\theta | \boldsymbol\theta^{(s)}) = - \sum_{k=1}^K \int \log [f(\boldsymbol y_k | \boldsymbol X_k, \boldsymbol\delta_k; \boldsymbol\theta)] \phi(\boldsymbol\delta_k | \boldsymbol d_{k,o}; \boldsymbol\theta^{(s)}) d \boldsymbol\delta_k,
  \label{eqn:Q1_old}
\end{equation}
where $\boldsymbol \theta$ includes the fixed effects $\boldsymbol \beta$, the non-zero elements of $\boldsymbol \Gamma$ given in (\ref{eqn:chp4_linpredgam}), and $\tau$, and the $\boldsymbol \delta_k$ are $q$-dimensional latent variables. However, by using the novel model formulation given in (\ref{eqn:chp4_linpredB}), this changes the integral of interest such that now $Q_1(\boldsymbol \theta | \boldsymbol \theta^{(s)})$ expressed in (\ref{eqn:chp4_Q1}) is of dimension $r \ll q$. This significantly reduces the computational complexity of estimating this integral in the E-step of the algorithm since we only have to estimate a latent space of dimension $r$. Consequently, this reduces the computational time. This also enables us to scale our method to hundreds of predictors since the practical dimension of the latent space will be much smaller than the total number of candidate random effects predictors. 
See Web Appendix Section 1.6 for further discussion about the values of $p$, $q$, and $r$ used in this paper.

Furthermore, this proposed formulation allows for more complex correlation structures in higher dimensions. In \cite{rashid2020}, the authors approximated the random effect covariance matrix $\boldsymbol \Sigma$ as a diagonal matrix when the dimensions are large as recommended by Fan and Li \citeyearpar{fanli2012}. This approximation reduced the computational complexity of the algorithm and therefore increased the speed of the model fit.
However, in our new formulation, we do not need to assume $\boldsymbol \Sigma$ is a diagonal matrix when the dimension is high.

\subsection{Estimation of the number of latent factors}
\label{sec:lit_r_est}

Performing our proposed \textbf{glmmPen\_FA} method requires specifying the number of latent factors $r$. Since $r$ is typically unknown \textit{a priori}, this value needs to be estimated. We estimate $r$ at the very beginning of the algorithm, and then use this estimated value in all later model estimations during the variable selection procedure.

There have been several proposed methods of estimating $r$ for the approximate factor model. 
We tried the Eigenvalue Ratio method and Growth Ratio method developed by \cite{ahn2013eigenvalue} as well as the  method proposed in \cite{bai_ng2002}. We found that the Growth Ratio (GR) method gave the most accurate estimates of $r$ within our application. Therefore, in this section, we will describe how we implement the GR method to estimate $r$. The GR method is used in all of our numerical works. 

To apply the GR method to our problem, we need a $q \times K$ matrix of observed random effects. Since we can never observe the random effects, we instead calculate pseudo random effects by first fitting a penalized generalized linear model with a small penalty to each group separately. We then take these group-specific estimates and center them so that all features have a mean of 0. Let these $q$-dimensional group-specific estimates be denoted as $\boldsymbol{\hat \gamma_k}$ for each group $k=1,...,K$. We then define $\boldsymbol G = (\boldsymbol{\hat \gamma_1},...,\boldsymbol{\hat \gamma_K})$ as the final $q \times K$ matrix of pseudo random effects. 

Let $\psi_j(A)$ be the $j$-th largest eigenvalue of the positive semidefinite matrix $A$, and let $\tilde \mu_{qK,j} \equiv \psi_j (\bx G \bx G^T / (qK)) = \psi_j (\bx G^T \bx G / (qK))$.

To find the GR estimator, we first order the eigenvalues of $\bx G \bx G^T / (qK)$ from largest to smallest. 
Then, we calculate the following ratios:
\begin{equation}
    GR(j) \equiv \frac{\log [V(j-1) / V(j)]}{\log [V(j)/V(j+1)]} = \frac{\log (1 + \tilde \mu_{qK,j}^*)}{\log (1 + \tilde \mu_{qK,j+1}^*)}, 
\end{equation}
for $j = 1,2,...,U$, where $V(j) = \sum_{l=j+1}^{\min(q,K)} \tilde \mu_{qK,l}$, $\tilde \mu_{qK,j}^* = \tilde \mu_{qK,j} / V(j)$, and $U$ is a pre-defined constant. Then, we estimate $r$ by
\begin{equation}
    \widehat{r}_{GR} = \text{max}_{1 \leq j \leq U} GR(j).
\end{equation}

\section{Simulations: Variable selection in Binomial data with 100 predictors}
\label{sec:sims}

We examine the variable selection performance of the \textbf{glmmPen\_FA} algorithm in high dimensions of $p=100$ total predictors under several different conditions of the underlying data. In all of these simulations, we use a pre-screening step to remove some random effects at the start of the algorithm in order to help improve the speed of the algorithm (see Web Appendix Section 1.2 for details), the BIC-ICQ \citep{BICq2011} criterion for tuning parameter selection, the MCP penalty (MCP penalty for the fixed effects, group MCP penalty for the rows of the $\boldsymbol B$ matrix), and the abbreviated two-stage grid search as described in the Web Appendix Section 1.2 using the penalty sequences described in Web Appendix Section 1.3. In order to determine the robustness of our variable selection procedure based on the assumed value of $r$, we fit models in one of two ways: we estimated the number of common factors $r$ using the Growth Ratio estimation procedure discussed in Section \ref{sec:lit_r_est}, or we input the true value of $r$ for the algorithm to use. All simulation conditions used 100 replicates.


We simulated binary responses from a logistic mixed effects model with $p=100$ predictors. Of \(p\) total predictors, we assume that the first 10 predictors have truly non-zero fixed and random effects, and the other \(p-10\) predictors have zero-valued fixed and random effects. We specified a full model as input for the algorithm such that the candidate random effect predictors equalled the candidate fixed effect predictors (i.e. assumed $q=p$), and our aim was to select the set of true predictors and random effects.

We set the sample size to $N=2500$ and number of groups to $K=25$, with an equal number of subjects per group. We set up the random effects covariance matrix by specifying a $\boldsymbol B$ matrix with dimensions $(p+1)\times r$, where $p+1$ represents the $p$ predictors plus the random intercept, and $r$ represents the number of latent common factors with $r=\{3,5\}$. Eleven of these $p+1$ rows---corresponding to the true 10 predictors plus the intercept---had non-zero elements, while the remaining $p-10$ rows were set to zero. For each value of $r$, we considered $\boldsymbol B$ matrices that produced covariance matrices $\boldsymbol \Sigma = \boldsymbol B \boldsymbol B^T$ with moderate variances and eigenvalues and large variances and eigenvalues (see Web Appendix Section 2.1 for further details). These $\boldsymbol B$ matrices are referred to as the `moderate' and `large' $\boldsymbol B$ matrices, respectively. We use both moderate predictor effects and strong predictor effects, where all 10 of the true fixed effects have coefficient values of 1 or 2, respectively.

For group \(k\), we generated the binary response \(y_{ki}\), \(i=1,...,n_k\) such that \(y_{ki} \sim Bernoulli(p_{ki})\) where \(p_{ki} = P(y_{ki} = 1 | \boldsymbol x_{ki}, \boldsymbol z_{ki}, \boldsymbol \gamma_k, \boldsymbol \theta) = \exp(\boldsymbol x_{ki}^T \boldsymbol \beta + \boldsymbol z_{ki}^T \boldsymbol \gamma_k) / \{1 + \exp(\boldsymbol x_{ki}^T \boldsymbol \beta + \boldsymbol z_{ki}^T \boldsymbol \gamma_k) \}\), and \(\boldsymbol \gamma_k \sim N_{11}(0, \boldsymbol B \boldsymbol B^T)\). Each condition was evaluated using 100 total simulated datasets.

For individual \(i\) in group \(k\), the vector of predictors for the fixed effects was \(\boldsymbol x_{ki} = (1,x_{ki,1},...,x_{ki,p})^T\), and we set the random effects \(\boldsymbol z_{ki} = \boldsymbol x_{ki}\), where \(x_{ki,j} \sim N(0,1)\) for \(j=1,...,p\), and each $\boldsymbol x_j$ was standardized as described in Section \ref{sec:models}.


The results for these simulations are presented in Tables \ref{tab:bin_pos} and \ref{tab:bin_r}. Table \ref{tab:bin_pos} provides the average true positive rates (percent of true predictors selected) and false positive rates (percent of false predictors selected) for both the fixed and random effects variable selection, the median time in hours to complete the variable selection procedure, the average of the mean absolute deviation between the fixed effect coefficient estimates and the true coefficients across all simulation replicates, and the average of the Frobenius norm of the difference between the estimated random effect covariance matrix $\boldsymbol{\hat \Sigma} = \boldsymbol{\hat B \hat B^T}$ and the true covariance matrix $\boldsymbol{\Sigma} = \boldsymbol{B B^T}$ (the Frobenius norm was standardized by the number of random effects selected in the best model). Table \ref{tab:bin_r} gives the Growth Ratio $r$ estimation procedure results, including the average estimate of $r$ and the proportion of times that the Growth Ratio estimate of $r$ was underestimated, correct, or overestimated. All simulations were completed on a Longleaf computing cluster (CPU Intel processors between 2.3Ghz and 2.5GHz).

\begin{table}[h!] 
\centering
\begin{tabular}{ccccccccccc}
  \hline
   True $r$ & $\beta$ & $\boldsymbol B$ & $r$ Est. & 
   \multicolumn{1}{p{1.25cm}}{\centering TP \% \\ Fixef} & 
   \multicolumn{1}{p{1.25cm}}{\centering FP \% \\ Fixef} & 
   \multicolumn{1}{p{1.25cm}}{\centering TP \% \\ Ranef} & 
   \multicolumn{1}{p{1.25cm}}{\centering FP \% \\ Ranef} & $T^{med}$ &
   \multicolumn{1}{p{1.25cm}}{\centering Abs. Dev. \\ (Mean)} &
   \multicolumn{1}{p{1.25cm}}{\centering $||\boldsymbol D||_F$ \\   }\\ 
  \hline
  3 & 1 & Mod. & GR      & 98.50 & 2.00 & 97.20 & 0.22 & 2.05 & 0.26 & 0.93 \\ 
    &   &       & True    & 99.00 & 2.14 & 98.40 & 0.16 & 2.36 & 0.26 & 0.93\\ 
    &   & Large & GR      & 95.50 & 2.19 & 98.60 & 0.18 & 2.52 & 0.33 & 1.94 \\ 
    &   &       & True    & 95.50 & 2.31 & 98.90 & 0.17 & 2.42 & 0.33 & 1.96 \\ 
    & 2 & Mod. & GR      & 100.00 & 2.46 & 89.00 & 0.53 & 1.45 & 0.37 & 0.89 \\ 
    &   &       & True    & 100.00 & 2.78 & 90.10 & 0.50 & 2.07 & 0.31 & 0.92 \\ 
    &   & Large & GR      & 100.00 & 3.39 & 94.60 & 0.80 & 2.39 & 0.43 & 1.78 \\ 
    &   &       & True    & 100.00 & 3.40 & 96.20 & 0.49 & 2.41 & 0.41 & 1.60 \\
  
  \hline
    5 & 1  & Mod. & GR      & 96.80 & 2.02 & 96.20 & 0.04 & 3.56 & 0.35 & 1.54 \\ 
      &   &       & True    & 96.70 & 1.86 & 96.80 & 0.03 & 3.60 & 0.35 & 1.59\\ 
      &   & Large & GR      & 90.40 & 2.22 & 96.80 & 0.08 & 4.39 & 0.44 & 2.73 \\ 
      &   &       & True    & 90.50 & 1.97 & 96.90 & 0.07 & 4.44 & 0.44 & 2.83 \\ 
      & 2  & Mod. & GR      & 100.00 & 2.11 & 89.00 & 0.18 & 2.29 & 0.52 & 1.22\\ 
      &   &       & True    & 100.00 & 2.42 & 88.40 & 0.24 & 2.99 & 0.44 & 1.33 \\ 
      &   & Large & GR      & 99.90 & 3.28 & 93.10 & 0.50 & 3.03 & 0.57 & 2.26 \\ 
      &   &       & True    & 99.90 & 3.36 & 93.40 & 0.47 & 3.98 & 0.55 & 2.29\\ 
   \hline \\
\end{tabular}
\caption{Variable selection results for the $p=100$ logistic mixed effects simulations, including true positive (TP) percentages for fixed and random effects, false positive (FP) percentages for fixed and random effects, the median time in hours for the algorithm to complete ($T^{med}$), and the average of the mean absolute deviation (Abs. Dev. (Mean)) between the fixed effect coefficient estimates $\hat \beta$ and the true $\beta$ values across all simulation replicates. Column $\boldsymbol B$ describes the general size of both the variances and eigenvalues of the resulting $\boldsymbol \Sigma = \boldsymbol{B B}^T$ random effects covariance matrix (moderate vs large). Column `$r$ Est.' refers to the method used to specify $r$ in the algorithm: the Growth Ratio estimate or the true value of $r$. Column $||\boldsymbol D||_F$ represents the average across simulation replicates of the Frobenius norm of the difference ($\boldsymbol{D}$) between the estimated random effects covariance matrix $\boldsymbol{\hat \Sigma}$ and the true random effects covariance matrix $\boldsymbol{\Sigma}$; the Frobenius norm was standardized by the number of true random effects selected in the model.}
\label{tab:bin_pos}
\end{table}

\begin{table}[h!] 
\centering
\begin{tabular}{ccccccc}
  \hline
   True $r$ & $\beta$ & $\boldsymbol B$ & Avg. $r$ & $r$ Underestimated \% & $r$ Correct \% & $r$ Overestimated \% \\ 
  \hline
  3 & 1 & Mod. & 2.79 & 21 & 79 & 0 \\ 
    &   & Large & 2.95 & 5 & 95 & 0 \\ 
    & 2 & Mod. & 2.21 & 80 & 19 & 1 \\
    &   & Large & 2.51 & 49 & 51 & 0 \\ 
  \hline
  5 & 1 & Mod. & 4.60 & 26 & 72 & 2 \\
    &   & Large & 4.83 & 15 & 83 & 2 \\ 
    & 2 & Mod. & 3.83 & 70 & 28 & 2 \\ 
    &   & Large & 4.43 & 46 & 50 & 4 \\ 
   \hline \\
\end{tabular}
\caption{Results of the Growth Ratio $r$ estimation procedure for $p=100$ logistic mixed effects simulation results, including the average estimate of $r$ across simulations and percent of times that the estimation procedure underestimated $r$, gave the true $r$, or overestimated $r$. Column $\boldsymbol B$ describes the general size of both the variances and eigenvalues of the resulting $\boldsymbol \Sigma = \boldsymbol{B B}^T$ random effects covariance matrix (moderate vs large).}
\label{tab:bin_r}
\end{table}

We see from Table \ref{tab:bin_pos} that the \textbf{glmmPen\_FA} method is able to accurately select both the fixed and random effects across a variety of conditions, which is supported by the true positives generally being above 90\% for both the fixed and random effects and the false positives generally being small: across all conditions, less than 3.5\% for fixed effects and less than 1\% for random effects.


We can see from Table \ref{tab:bin_r} that the Growth Ratio (GR) estimation procedure applied to the pseudo random effect estimates described in Section \ref{sec:lit_r_est} has varying levels of accuracy depending on the structure of the underlying data. Generally, the GR estimation procedure becomes more accurate as the eigenvalues of the covariance matrix increase and the true predictor effects are moderate. 
We have found that the estimation of $r$ generally improves when the sample size per group increases (simulations not shown) or when the total number of predictors used in the GR estimation procedure decreases (compare Table \ref{tab:bin_r} with Web Appendix Sections 2.2 and 2.3).
Under conditions that reduce the accuracy of the GR procedure, the GR procedure underestimates $r$ on average.
However, when we compare the true and false positive rates for the fixed and random effects given in Table \ref{tab:bin_pos} between scenarios using the true $r$ and those using the estimated $r$, we see very similar results, even in situations when the GR procedure tended to underestimated $r$. The mis-specification of $r$ does not significantly impact the estimation of the fixed effects coefficients (see the mean absolute deviation values) nor does it significantly impact the estimation of the random effect covariance matrix coefficients (see the Frobenius norm values).

We also used \textbf{glmmPen} to perform variable selection on the simulations where the true number of latent factors was $r=3$. 
We let the \textbf{glmmPen} variable selection procedure proceed for 100 hours. In that time, \textbf{glmmPen} was able to complete the following number of replicates out of the 100 total replicates: 83 for ($\beta=1,\boldsymbol B=$ Moderate), 71 for ($\beta=1,\boldsymbol B= $ Large), 100 for ($\beta=2,\boldsymbol B=$ Moderate), and 96 for ($\beta=2,\boldsymbol B=$ Large). The minimum times needed to complete the \textbf{glmmPen} variable selection procedures were 39.91, 57.60, 23.63, and 42.79 hours, respectively.

The Web Appendix Section 2 contains additional simulation results not included in the main paper due to space considerations. These additional simulations include simulations with $p=500$ predictors, a comparison with \textbf{glmmPen} in moderate dimensions, and alternative data set-ups (e.g. sample size, effect magnitudes, correlated predictors).

\section{Case study: Pancreatic Ductal Adenocarcinoma}
\label{sec:PDAC}

Patients diagnosed with Pancreatic Ductal Adenocarcinoma (PDAC) generally face a very poor prognosis, where the 5-year survival rate is 6\% \citep{khorana2016potentially}. 
The study by \cite{moffitt2015} identified genes that are expressed exclusively in pancreatic tumor cells. Using these tumor-specific genes, \cite{moffitt2015} was able to identify and validate two novel tumor subtypes, termed `basal-like' and `classical'. It was found that patients diagnosed with basal-like tumors had significantly worse median survival than those diagnosed with the classical tumors. Consequently, it is of clinical interest to robustly predict this basal-like subtype in order to make and improve tailored treatment recommendations. 


In order to improve replicability in the prediction of subtypes in PDAC, we combine PDAC gene expression data from five different studies \citep{aguirre2018real,cao2021proteogenomic,dijk2020unsupervised,hayashi2020unifying,raphael2017integrated} with a total sample size of 360 subjects; see Web Appendix Table 9 for further details.
In order to account and adjust for between-study heterogeneity, we apply our new method \textbf{glmmPen\_FA} to fit a penalized logistic mixed effects model to our data to select predictors with study-replicable effects, where we assume that predictor effects may vary between studies. 

The basal or classical subtype outcome was calculated using the clustering algorithm specified in \cite{moffitt2015}. Further details are provided in Web Appendix Section 3.1, and the code for this procedure is provided in a GitHub repository, see Supplementary Materials for more details.

\cite{moffitt2015} identified a list of 500 genes that were likely to be expressed solely in PDAC tumor cells. 
The five studies had RNA-seq gene expression data for 432 of these 500 genes. There were some significant correlations between some of these 432 genes, as evaluated by Spearman correlations applied to the subject-level rank-transformed gene expression; therefore, we combined highly correlated genes together into meta-genes.
The clustering process used to create these meta-genes is described in Web Appendix Section 3.1. The final dataset included 117 meta-genes. 
The raw gene expression values of each meta-gene, calculated as the sum of the gene expression across all of the genes comprising the meta-gene, were converted from their raw values to their ranks for each subject.
 
Due to the presence of several pairwise Spearman correlation values greater than 0.5 in this final dataset, we used the Elastic Net penalization procedure \citep{glmnet2010} to balance between ridge regression and the MCP penalty. We let $\pi$ represent the balance between ridge regression and the MCP penalty, where $\pi=0$ represents ridge regression and $\pi=1$ represents the MCP penalty. We restricted our consideration to $\pi=\{0.6,0.7,0.8,0.9\}$ based on Elastic Net simulation results given in Web Appendix Section 2.4; we then used sensitivity analyses (see Web Appendix Section 3.2) to choose the optimal $\pi$ used in this case study variable selection analysis.
Within each value of $\pi$, we considered the Growth Ratio estimated value of $r$ (evaluated at 2 for all values of $\pi$) and a manually set larger value of $r=3$. We found that within values of $\pi$, the selection results and coefficient values were consistent for the different values of $r$ considered; we therefore restricted our consideration to the Growth Ratio estimate of $r$. We also found that the selected meta-genes within the final model 
were consistent across the values of $\pi$, with only small deviations between values of $\pi$. 
For the results reported here, we let $\pi=0.8$, which provided the results that best reflected the conclusions from the overall sensitivity analyses. The same value of $\pi$ was used for both the fixed effects and random effects penalization. The sequence of $\lambda$ penalties used in the variable selection procedure was the same as those used in the Binomial variable selection simulations for $p=100$ (see Web Appendix Section 1.3).


In the final results, 8 of the 117 total meta-gene covariates had non-zero fixed effect values in the best model selected by the BIC-ICQ criteria, implying these covariates were important for the prediction of the basal outcome. These 8 meta-gene covariates represented 37 genes in total. Table \ref{tab:PDAC_results} includes the label for these 8 meta-genes, the sign of the associated fixed effect coefficient (i.e. the log odds ratio estimate), and the gene symbols of the genes that make up the meta-gene. Meta-genes with positive log odds ratios indicate that having greater relative expression of these meta-genes increases the odds of a subject being in the basal subtype, and vice versa for negative log odds ratios. The best model contained a random intercept (variance value 0.54) and no other random slopes. 

\begin{table}[h!] 
    \centering
    \begin{tabular}{ccl}
    \hline
    \multicolumn{1}{p{2cm}}{\centering Meta-gene \\ No.} & 
    \multicolumn{1}{p{2.5cm}}{\centering Log Odds \\ Ratio Sign} & Gene Component Symbols for Meta-Gene \\
   \hline
    5 & + & ADORA2B, C16orf74, HES2, ULBP2\\
    7 & + & AHNAK2, FAM83A, GJB6, ITGA3, IVL, KRT6A,\\
      &   &  MUC16, PPL, SCEL, SLC2A1, ZNF185\\
    28 & + & COL17A1, DHRS9, SPRR1B, SPRR3\\
    52 & + & KRT23, S100A4, SPINT2\\
    81 & + & BACE2, CEACAM3, RNF43\\
    85 & - & BTNL8, CDH17, MYO1A, MYO7B, PDZD3, VIL1\\
    104 & - & GATA6, PAQR8, PIP5K1B, TOX3\\
    117 & - & TFF1, VSIG2 \\
    \hline \\
    \end{tabular}
    \caption{Covariate meta-gene label within the case study dataset of the meta-genes that had non-zero fixed effects in the final best model, the sign of the fixed effect coefficient (i.e. the sign of the log odds ratio) associated with the meta-gene, and the gene symbols of the genes within the meta-gene.}
    \label{tab:PDAC_results}
\end{table}


We also applied the \textbf{glmmPen} variable selection procedure to this data. Using $\pi=0.8$, the 8 meta-gene covariates selected by \textbf{glmmPen\_FA} were also selected by \textbf{glmmPen}. The \textbf{glmmPen} method selected two additional meta-genes (meta-gene 59, genes PKIB, DNAJC15, with negative log odds ratio; meta-gene 71, genes AKR1C3, CA2, MGST2, with positive log odds ratio) and selected meta-gene 117 to have a non-zero random effect (variance 0.99). The main difference in these variable selection procedures was the time to needed to complete the procedure, where \textbf{glmmPen\_FA} finished within 0.8 hours and \textbf{glmmPen} finished within 49.4 hours. More details about \textbf{glmmPen} sensitivity analyses (i.e. results for different values of $\pi$) are provided in Web Appendix Section 3.2.

\section{Discussion}
\label{sec:discuss}

By adopting a factor model structure to estimate the high-dimensional random effect covariance matrix in the generalized linear mixed model setting, we are assuming that a small number of underlying latent variables (i.e. latent factors) can fully describe the high dimensional set of candidate random effects we consider in the model. The main benefit of this assumption is that we are able to reduce the latent space from a large number of random effects to a smaller set of latent factors, thereby greatly simplifying the Expectation step (E-step) of the algorithm. We have shown through simulations (both in Section 3 and in the supplementary simulations in Web Appendix Section 2) that by reducing the complexity of the integral in the E-step, we can significantly improve the overall time needed to perform variable selection in high dimensional generalized linear mixed models. 

The simulations 
also show how reducing the latent space increases the feasible dimensionality of performing variable selection in generalized linear mixed models. By using our novel formulation of the random effects, we can perform variable selection on mixed models with hundreds of predictors within a reasonable time-frame without any \textit{a priori} knowledge of which predictors are relevant for the model, either in terms of fixed or random effects. From the simulation results, we see that the \textbf{glmmPen\_FA} method results in accurate selection of the fixed and random effects across several conditions. 

One limitation of this assumption is that there may be applications where the random effects cannot be represented as a function of a relatively small set of latent factors. Our method provides the greatest computational benefit (i.e. the greatest improvement in time) when the true value of the number of latent factors $r$ is much smaller than the number of random effects considered by the variable selection procedure, $q$. If the value of $r$ from the estimation procedure is large, or not much smaller than the number of random effects $q$, then our method has little computational advantage over the existing method of \textbf{glmmPen}.



Additionally, our method is limited by the need to provide an estimate for the number of latent common factors. 
However, the simulation results show that our data-driven estimation of the number of latent factors, based on the Growth Ratio estimation procedure by \cite{ahn2013eigenvalue}, provides reasonable estimates. Even when it was estimated incorrectly by this procedure, this mis-specification had very little impact on the general variable selection performance or the coefficient estimates. Therefore, our method is not sensitive to the estimation of the number of latent factors. 




\backmatter


\section*{Acknowledgements}

Research reported in this publication was supported the National Institutes of Health under the following award numbers: RO1 AG073259, U01 CA274298, P50 CA257911, P50 CA058223, and T32 CA106209.

\section*{Supplementary Materials}
Web Appendices and Tables referenced in Sections \ref{sec:methods}, \ref{sec:sims}, and \ref{sec:PDAC}, Section \ref{sec:sims} code and simulation output, and Section \ref{sec:PDAC} data and code are available with this paper at the Biometrics website on Oxford Academic and the GitHub repository \url{https://github.com/hheiling/paper_glmmPen_FA}. 
The \textbf{glmmPen} R package containing both the original \textbf{glmmPen} formulation and the new \textbf{glmmPen\_FA} method is available for download through CRAN at \url{https://cran.r-project.org/package=glmmPen}. 

\section*{Data Availability}
The data that support the findings in this paper are provided in the Supplementary Materials for this paper.

\bibliographystyle{biom} \bibliography{Biom_Biblio}

\label{lastpage}

\end{document}